\documentclass[prb, a4paper, preprint, fleqn, twoside, floatfix]{revtex4}
\pdfoutput=1
\usepackage{times}
\usepackage{graphicx}
\usepackage{amssymb}
\usepackage{wasysym}
\usepackage[usenames,dvipsnames]{color}

\newcommand{\m}[1]{{\mathrm{{#1}}}} %  remove math
\newcommand{\s}[1]{{_{\mathrm{{#1}}}}} % do subscript and remove math

\begin{document}
\received{\today}
\title{Diffusion of O$_2$ and N$_2$ through thin and thick SWNT networks}

\author{C Morgan}
\email{c.j.morgan@qmul.ac.uk}

\author{M Baxendale}

\affiliation{Molecular and Materials Physics Group, Department
of Physics, Queen Mary, University of London, Mile End Rd.,
London E1 4NS, UK}

\begin{abstract}
We report the electrical responses of thin and thick single-walled carbon nanotube (SWNT) networks to N$_2$ and O$_2$ adsorption. In the surface desorbed state exposure to N$_2$ and O$_2$ provide an increase in conductance of thin and thick SWNT networks. The increase in conductance of both thin and thick networks is of a greater magnitude during O$_2$ exposure rather than N$_2$ exposure.  Thin networks exhibit a greater response to both O$_2$ and N$_2$ rather than thick networks. This is likely a result of the increased semiconducting nature of thin SWNT networks.
\end{abstract}

\maketitle

\section{Introduction}
Since the discovery of carbon nanotubes (CNTs) in 1991 by Ijima \cite{iijima} activity among researchers has focused on exploring and exploiting the unique electrical, physical and mechanical properties of CNTs. Single walled carbon nanotube (SWNT) networks are increasingly used as the transduction layer in gas sensors and as storage media for gas storage devices \cite{Kauffman2008a,Baughman2002}.

Recent application of SWNT networks as the gas diffusion layer in fuel cells \cite{Kaempgen2008} and in hydrogen storage devices \cite{Benard2007} require a thorough understanding of the diffusion of gasses through SWNT networks and the influencing factors in order to optimise the device performance.

Adsorbed gases significantly alter the transport properties of SWNTs \cite{Bradley2000,Collins2000,Morgan2008,Lee2008a}. Analysis of the temporal response of conductance as molecular adsorption occurs through the network provides information about the interaction between gas molecules and SWNT networks. Zahab \emph{et al.} \cite{Zahab2000} observed gaussian diffusion to occur for water adsorption through a SWNT network at two different diffusion rates. This study investigates the diffusion of oxygen and nitrogen through thick and thin SWNT networks. Analysis of the temporal response of conductance in terms of diffusion phenomena suggests distinct regions of diffusion, characterised by the diffusion parameter $d$. At later times the temporal change in conductance displays an activated behaviour suggesting either pre-existing surface heterogeneity or adsorption induced surface heterogeneity.

\section{Experimental details}

SWNTs produced via the high-pressure carbon monoxide (HiPco) method were purchased from Carbon Nanotechnologies Incorporated (Texas). An aqueous solution containing 0.44 wt$\%$ SWNT with 1 wt$\%$ SDS (Sigma-Aldrich, Product No. L6026) was prepared. The SWNT/SDS solution was subject to mild probe sonication of power 100 W and frequency 20 kHz for 1 hour 30 mins. Centrifugation at 3000 U/min for 2 hours separated higher density contaminants from dispersed SWNTs, the SWNT supernatant was extracted.  The solution was then diluted by a factor of 100 before deposition.

\begin{figure}
\centering
\includegraphics[width=0.6\textheight]{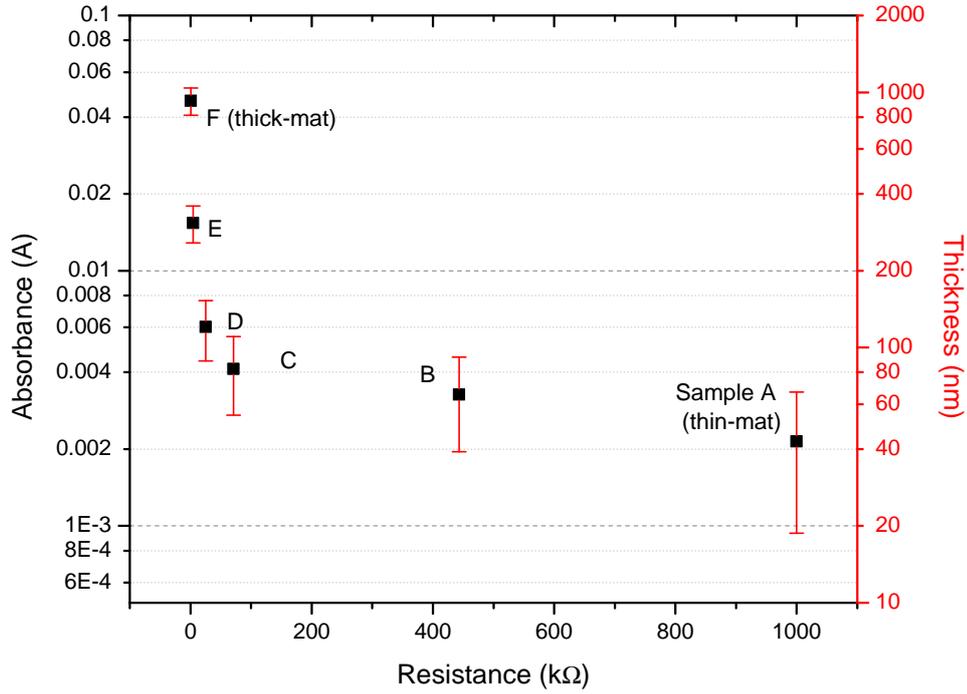}
\caption{Resistance versus absorbance at $\lambda = 500$ nm and thickness of SWNT network. Error bars represent error in Beer's law derived SWNT network thickness.}
\label{ResSen:NormTvsR}
\end{figure}

Samples of increasing thickness were deposited on quartz substrates with two single stripe metal contacts of length 1cm, width 250 $\mu$m and separation 0.4 cm. UV-visible (UV-vis) measurements were performed with a Perkin Elmer Lambda 950. Previous studies \cite{Bahr2001,Zhao2004} have confirmed the validity of Beer's law for SWNT solutions, providing SWNT extinction coefficients at $\lambda = 500$ nm in the range $2.86 \times10^{4} \ \m{cm}^2\m{g}^{-1}$ $< \epsilon_{500 \ \m{nm}} < 3.5 \times 10^{4} \ \m{cm}^{2} \m{g}^{-1}$ respectively. The variation of $\epsilon$ in the above studies can be attributed to the fraction of individual SWNT versus SWNT bundles present in the measurement sample as a result of differing filtration methods and differing organic solvent used to form the dispersions. Nanotube bundles may scatter incident light and thus be the cause of interference in UV-vis absorption measurements \cite{Zhou2003}. The UV-vis absorption spectrum of bundled SWNT has been shown to be red-shifted and broadened \cite{smalley_band_flourese}, the origin of the spectral red-shift possibly resulting from a decrease in the band-gap of SWNT as a consequence of bundle formation \cite{iso_bundle}. The two-terminal resistance of the SWNT networks were measured and correlated with the absorbance of the network at $\lambda =  500$ nm and the thickness of the films calculated through Beer's law, Figure \ref{ResSen:NormTvsR}. This study is concerned with the properties of the thinnest (Sample A) and thickest (Sample F) SWNT networks from the sample set, henceforth referred to as thin and thick respectively.

The determination of SWNT network thickness through SWNT absorbance measurements and Beer's law has also been investigated by Bekyarova \emph{et al.} \cite{Bekyarova2005} and confirmed through AFM measurements. The density of deposited nanotubes were found by Bekyarova \emph{et al.} to be in the range $1.57  \ \m{g}\cdot  \m{cm}^{-3}< d < 1.59 \ \m{g}\cdot \m{cm}^{-3}$ through a density gradient method \cite{Bekyarova2005}. The values of $\epsilon$ and $d$ reported in the literature were employed in calculating the upper and lower limits of SWNT network thickness through Beer's law. The variation in the values of $\epsilon$ and $d$ obtained from the literature give the error in $l = \pm 11 \%$, the final error in the thickness of SWNT networks include the relative error in the absorbance.

Raman spectra were recorded using microscale resolution with a Renishaw Raman System Model 3000 at excitation wavelengths $\lambda\s{exc} = 632.5$ nm and a Nicolet 910 FTIR at excitation wavelengths $\lambda\s{exc} = 532$ nm and $\lambda\s{exc} = 785$ nm. The mean average $G/D$ ratio of all three excitation wavelengths is $I\s{G}/I\s{D} = 9.7$.

\begin{figure}
  \centerline{
\mbox{\includegraphics[width=0.3\textwidth]{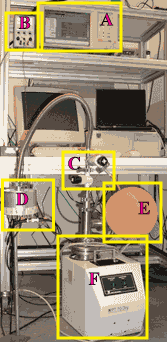}}}
\caption{Experimental set-up for sensing experiments.  A - Keithley 4200-SCU data acquisition system. B - UV-LED power supply. C - gas in valve and isolate pump valve. D - sample chamber. E -  test gas bladder. F - Leybold Dry-M-Basic oil free pump.}\label{ExpTech:gas_setup}
\end{figure}

The experimental setup utilised for gas sensitivity experiments is given in Figure \ref{ExpTech:gas_setup}.  The samples under test were mounted in a vacuum-tight sample box equipped with 12 mW UV LED, with a typical wavelength $\lambda\s{typical} = 400$ nm (RS Components, product code: 454-4396) situated 2 cm above sample and with an intensity of 0.03 mW/cm$^2$. The sample chamber is connected to an oil-free turbomolecular vacuum pump (Leybold, model: Dry-M-Basic) via two valves. One valve has an attachment for a bladder to facilitate back filling of gases into the sample chamber, providing exposure of the device to the analyte gas at atmospheric pressure. The second valve was used to isolate the turbomolecular vacuum pump. Oxygen and nitrogen gases were obtained from BOC and are of the following purity, nitrogen $99.998 \%$,  oxygen $99.5 \%$.

\begin{figure}
\centering
\includegraphics[width=0.5\textheight]{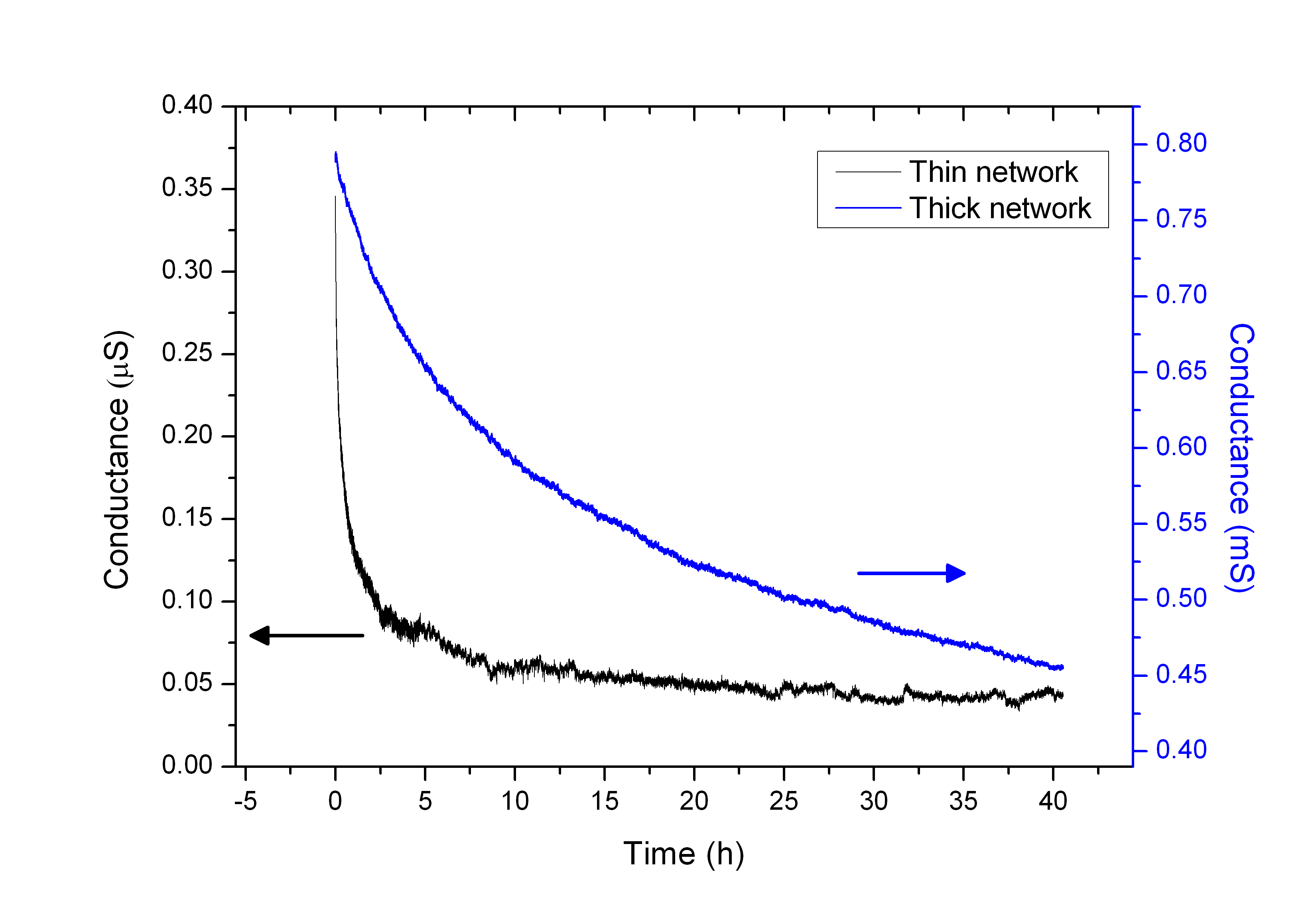}
\caption{Thin (\textbf{---$\!$---}) and thick ({\color{blue}{\textbf{---$\!$---}}}) SWNT networks exposed to UV radiation for 40 hours. Sample thickness determines the rate of desorption.}
\label{ResSen:SampleA_SampleF_UV}
\end{figure}

The conductance of the samples were monitored by periodically
sampling \mbox{($\Delta t = 1$ s)} the current whilst applying a fixed bias of $\leq 10$ mV, using a Keithley 4200-SCU data acquisition system.  Prior to adsorption experiments samples were placed under vacuum of \mbox{$\sim 1 \ \times 10^{-6} \ \mathrm{mbar}$} and exposed to UV LED illumination for 40 hours whilst measuring the conductance with time as described. This serves to desorb surface and inter-bundle adsorbates (surface dopants) from the nanotubes, evidenced in a conductance decrease. Figure \ref{ResSen:SampleA_SampleF_UV} details the change in conductance $\Delta G$ of air exposed thin and thick nanotube networks under UV-desorption. Once the conductance of the SWNT networks with time is stable, exposure to various analytes may be accomplished at atmospheric pressure.

\section{Results}

\begin{figure}
\centering
\includegraphics[width=0.5\textheight]{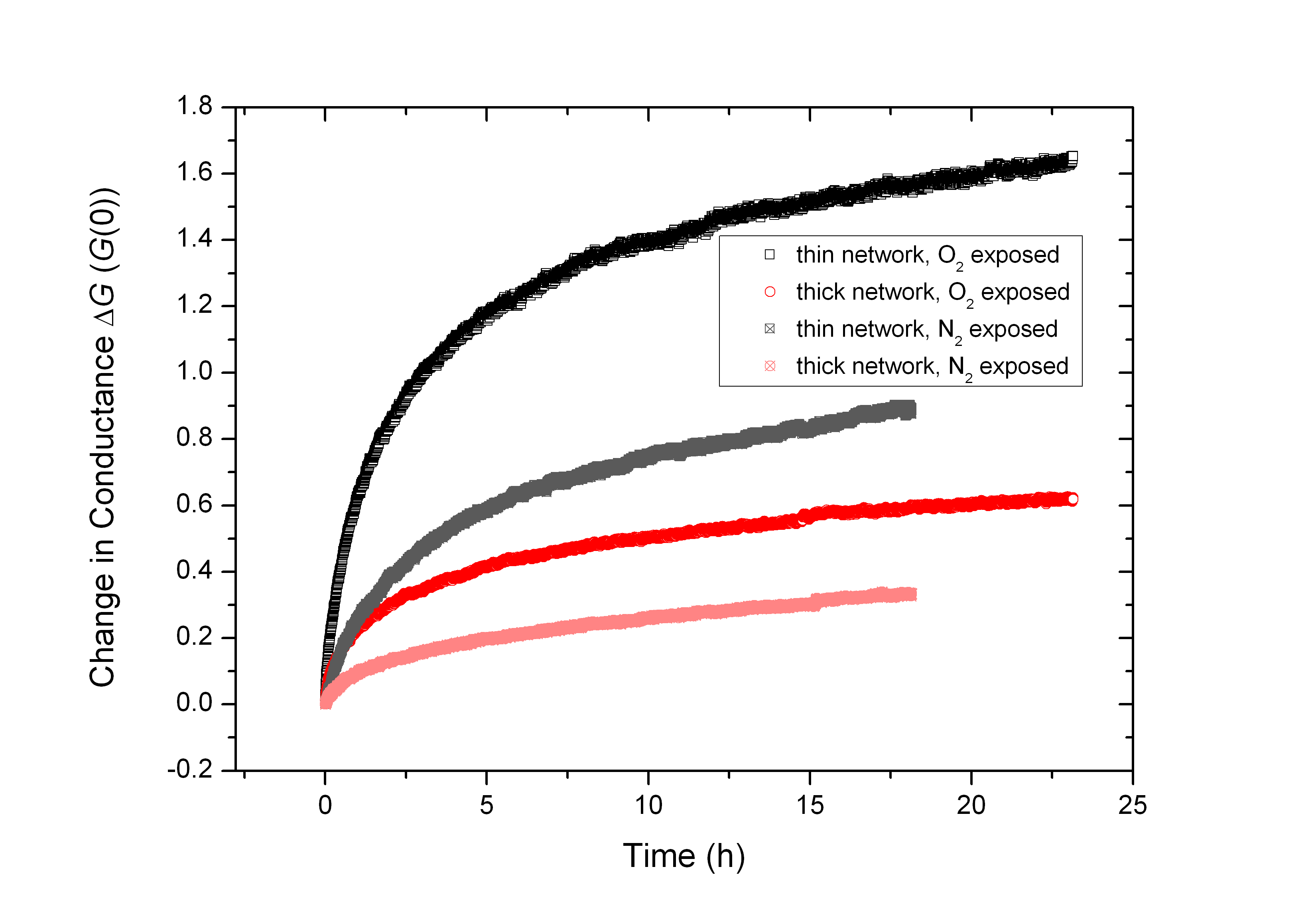}
\caption{Fractional change in conductance $\Delta G = G/G(0)-1$ versus time $t$ in hours for O$_2$ and N$_2$ adsorption for thin ($\square,{\color{Gray}{\boxtimes}}$) and thick (${\color{red}{\ocircle}},{\color{Salmon}{\otimes}}$) SWNT networks respectively.}
\label{increase}
\end{figure}

Figure \ref{increase} shows the fractional change in conductance $\Delta G$ of thin and thick SWNT networks exposed to atmospheric pressure oxygen and nitrogen from a UV-desorbed state. Saturation in $\Delta G$ occurs after $\sim 24$ hours.

The observed time dependence of the fractional change in conductance of thin and thick networks to both nitrogen and oxygen can be well described initially by a diffusive adsorption process.

\begin{figure}
\centering
\includegraphics[width=0.5\textheight]{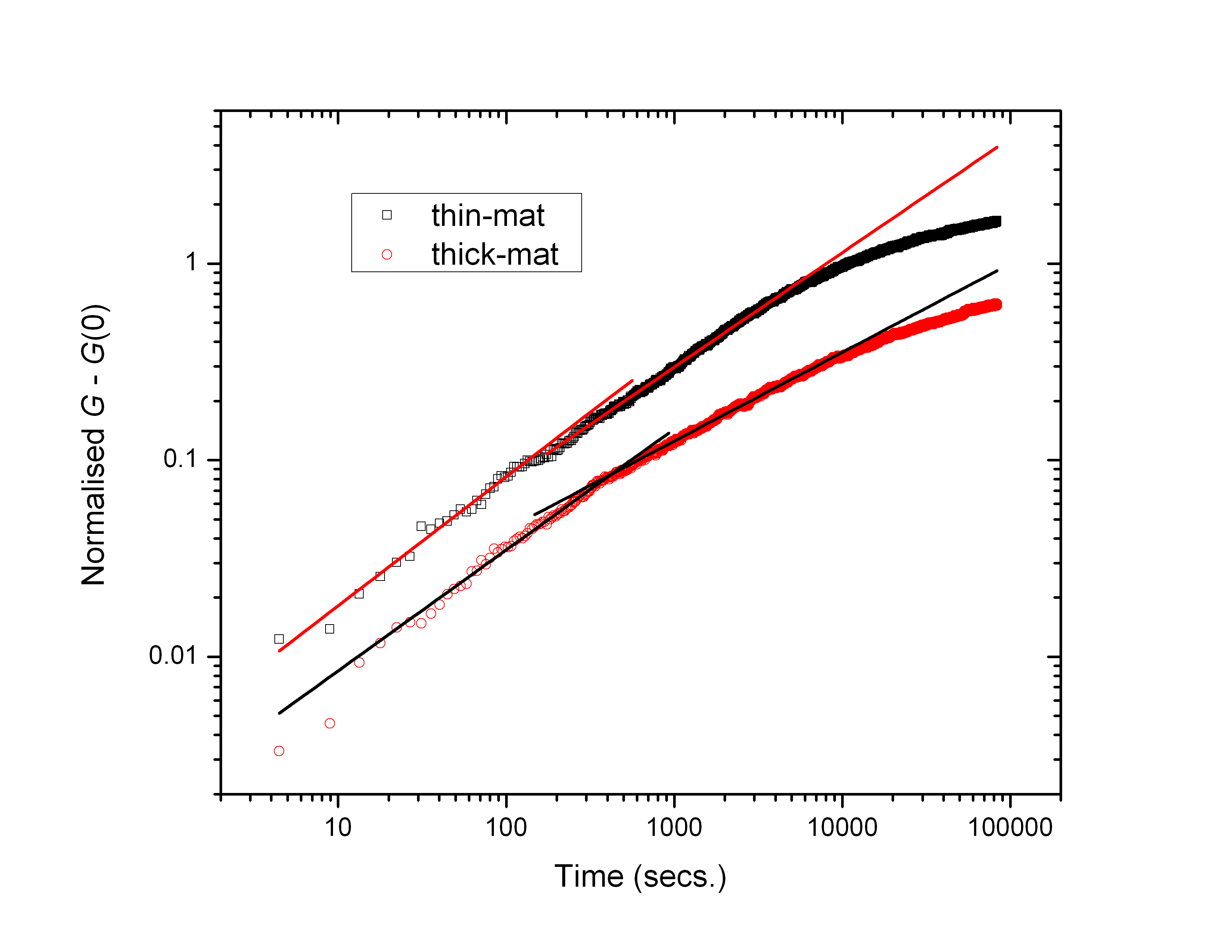}
\caption{Fractional change in conductance $\Delta G = G/G(0)-1$ versus time $t$ in hours for O$_2$ adsorption for thin ($\square$) and thick (${\color{red}{\ocircle}}$) SWNT networks respectively. Two linear regions are observed for thin network ({\color{red}{\textbf{---$\!$---}}}) and thick (\textbf{---$\!$---}) network respectively on a log-log axis scale.}
\label{o2_log_log}
\end{figure}

\begin{figure}
\centering
\includegraphics[width=0.5\textheight]{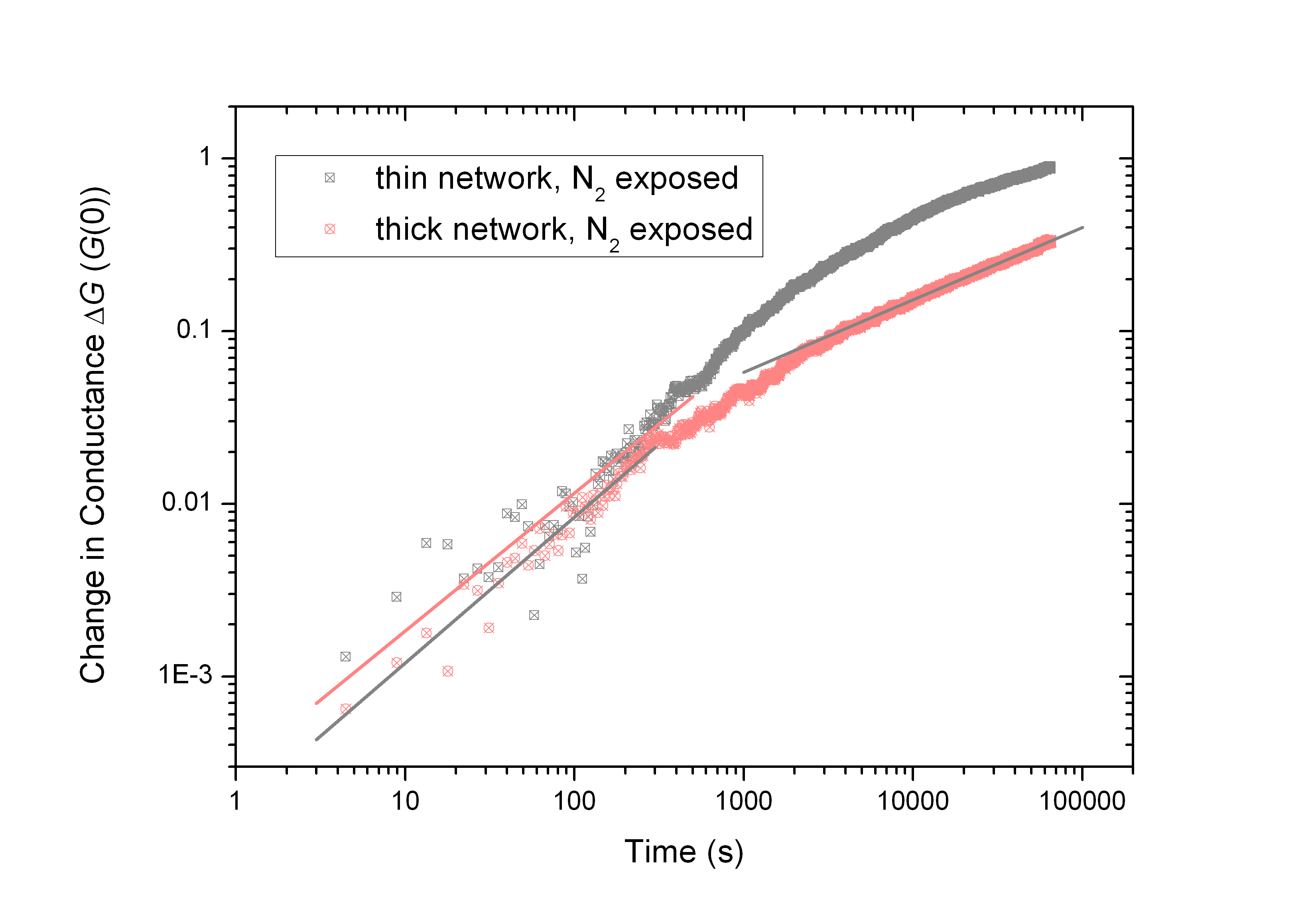}
\caption{Fractional change in conductance $\Delta G = G/G(0)-1$ versus time $t$ in hours for O$_2$ adsorption for thin (${\color{Gray}{\boxtimes}}$) and thick (${\color{Salmon}{\otimes}}$) SWNT networks respectively. Plot on log-log axis scale one linear region is observed for the thin network ({\color{Salmon}{\textbf{---$\!$---}}}) whilst two linear regions are observed for the thick network ({\color{Gray}{\textbf{---$\!$---}}}).}
\label{n2_log_log}
\end{figure}

The diffusion front in an isotropic material has a Gaussian profile

\begin{equation}
C(x,t) = \frac{s}{2\sqrt{\pi Dt}}\exp{(-x^2/4Dt)} \label{gaussian}
\end{equation}

\noindent where $s=\int_{-\inf}^{+\inf} Cdx$ the total amount of substance and $D$ is the molecular diffusion coefficient. As the concentration of gas molecules spreads out through the network a proportional amount of charge transfer occurs. The variation in conductance $\Delta G$ can be considered to be proportional to the standard deviation of the spatial distribution of gas molecules. The adsorption kinetics of thin and thick SWNT networks exhibit linear regions when plot on log-log axes, Figure \ref{o2_log_log} and \ref{n2_log_log}, indicating a relationship of the form

\begin{equation}
\Delta G = a\cdot t ^d \label{myvariance}
\end{equation}

\noindent where $d$ is the diffusion parameter and $a$ is a
constant. Two values of $d$ are observable when thin and thick networks are exposed to oxygen from a UV-desorbed state, Figure \ref{o2_log_log}.

Exposure to nitrogen from a UV-desorbed state results in an initial diffusive component for thin and thick networks. The thick network then displays a change in the value of $d$ which is maintained until the end of the experiment. The thin network on exposure to nitrogen departs from diffusive behaviour at approximately $t \sim 1000$ seconds, Figure \ref{n2_log_log}. Parameters of Equation \ref{myvariance} are extracted from Figure \ref{o2_log_log} and Figure \ref{n2_log_log} and given in Table \ref{SenCond:kine:DevAF:table1}.

The variance of a normal diffusive process is defined as

\begin{equation}
\sigma^2 = 2Dt^\gamma
\end{equation}

\noindent where $\sigma$ is the standard deviation of mass distribution and $\gamma$ is an index which describes the
diffusive regime ($\gamma = 1$ describes normal diffusion, $1 < \gamma < 2$ describes super-diffusion, $\gamma < 1$ describes sub-diffusion and $\gamma = 2$ describes ballistic diffusion \cite{levy,superdiff}). The variation in conductance $\Delta G$ can be considered to be proportional to the standard deviation of mass distribution through the network, resulting in $d = \gamma / 2$. Therefore, for a normal diffusive process $\sigma = 2Dt^{1/2}$ giving $d = 0.5$ in Equation \ref{myvariance}. Both thin and thick SWNT networks have $d_1 > 0.5$ in the initial time period of oxygen and nitrogen adsorption, Table \ref{SenCond:kine:DevAF:table1}, characteristic of super-diffusion \cite{superdiff} where $<x^2(t)>$ evolves non-linearly with time. Super-diffusive behaviour or enhanced diffusion is observed in porous systems, fractal geometries and Levy flights. During oxygen adsorption a change to a lower gradient occurs, $d_2$, for both thin and thick networks.  This second regime, characterised by $d_2$, was observed over a much longer time period, $\Delta t > 79$ minutes. In the case of nitrogen adsorption there is no well defined change in gradient.

\begin{table}\begin{center}
\begin{tabular}{lcccc}
    Sample & $d_1$ & Range of $t$ (s)& $d_2$ & Range of $t$ (s)\\
\hline \hline
thin O$_2$-adsorbed&      $0.66 \pm 0.02$  &    $0<t<174$ &      $0.582 \pm 0.001$ & $174<t<4900$ \\

thick O$_2$-adsorbed&      $0.58 \pm 0.02$ &    $0<t<433$ &     $ 0.451 \pm 0.001$& $433<t<5800$ \\

thin N$_2$-adsorbed & $0.80 \pm 0.04$ & $0<t<406$ & NA & NA \\

thick N$_2$-adsorbed & $0.85 \pm 0.02$ & $0<t<259$ & $0.420 \pm 0.001$ &$3729 <t<65334$ \\
\end{tabular}\end{center}
\caption{Extracted diffusion constants for oxygen and nitrogen adsorption from a UV-desorbed state, Figure \ref{o2_log_log} and Figure \ref{n2_log_log}.}
\label{SenCond:kine:DevAF:table1}
\end{table}

The thick SWNT network diffusion parameter $d_2 \sim 0.4$ in both the oxygen adsorbed and nitrogen adsorbed state and is characteristic of sub-diffusive behaviour. Abnormal diffusion of this type ($\gamma < 1$) is more common \label{Levy} and can be attributed to a trapping induced broad distribution of particle release time \cite{levy}.

Possible trapping sites within the nanotube network are SWNT defects \cite{defects}, semiconducting SWNT junctions
\cite{local_DOS} and nanotube bundle groves and interstitial sites \cite{Ulbricht2002}. Also, adsorption onto metallic-SWNT or bundles with metallic-SWNT at the surface may act as trapping sites in terms of doping efficiency (small shift of the  Fermi level provide small change in the density of states at the Fermi level). The lower value of $d$ for the thick network during oxygen adsorption could therefore be explained by an increase in the metallic component within the filamentary conduction path \cite{Morgan2008,Skakalova2006}. Zahab \emph{et al.} observed gaussian diffusion to occur for water adsorption on a SWNT mat of thickness 200 $\mu$m in which the SWNT material was purified by means of acid treatment and annealing under nitrogen atmosphere at 1200$^\circ$C \cite{Zahab2000}. Two regimes of characteristic diffusion parameter were found by Zahab to occur - a fast regime during the first 15 minutes with diffusion parameter $d = 0.7$ and a slower regime with characteristic diffusion parameter $d = 0.35$ occurring from 15 minutes to 14 hours. The difference in diffusion parameters and length of time over which they occur between the results reported here and those reported by Zahab \emph{et al.} reflects the variation in processing of the SWNT material, thickness of the SWNT mat and also the electronic and steric differences between water and O$_2$. Possibly the very long time over which Zahab \emph{et al.} observes diffusion to be characterised by sub-diffusive behaviour is a result of hydrogen bonding between water molecules inhibiting molecular diffusion.

\begin{figure}
\centering
\includegraphics[width=0.5\textheight]{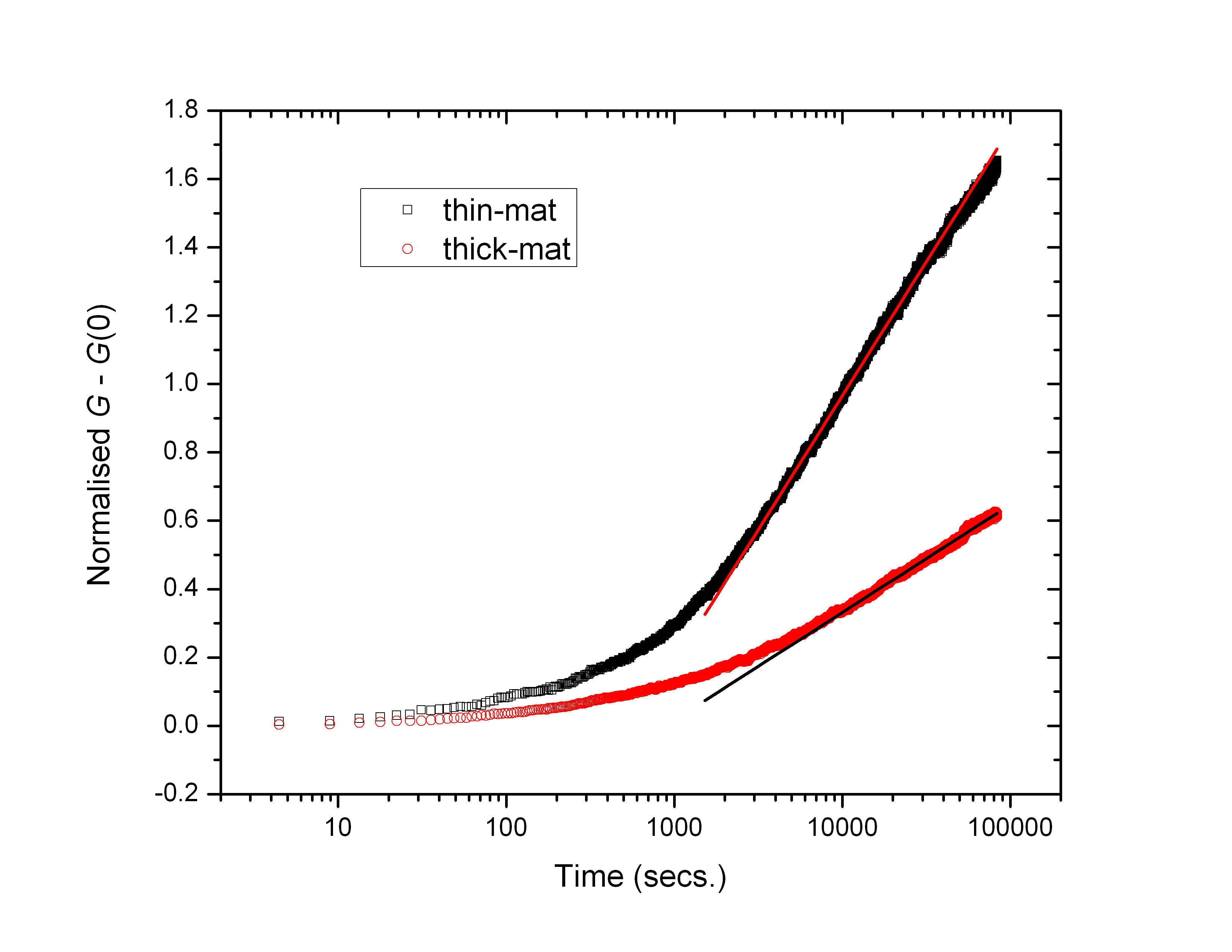}
\caption{Fractional change in conductance $\Delta G = G/G(0)-1$ versus time $t$ in hours for N$_2$ adsorption for thin ($\square$) and thick (${\color{red}{\ocircle}}$) SWNT networks respectively. Plot on linear-log axis scales one linear region is observed for the thin network ({\color{red}{\textbf{---$\!$---}}}) and thick (\textbf{---$\!$---}) network respectively.}
\label{SampleA_SampleF_logt}
\end{figure}

\begin{figure}
\centering
\includegraphics[width=0.5\textheight]{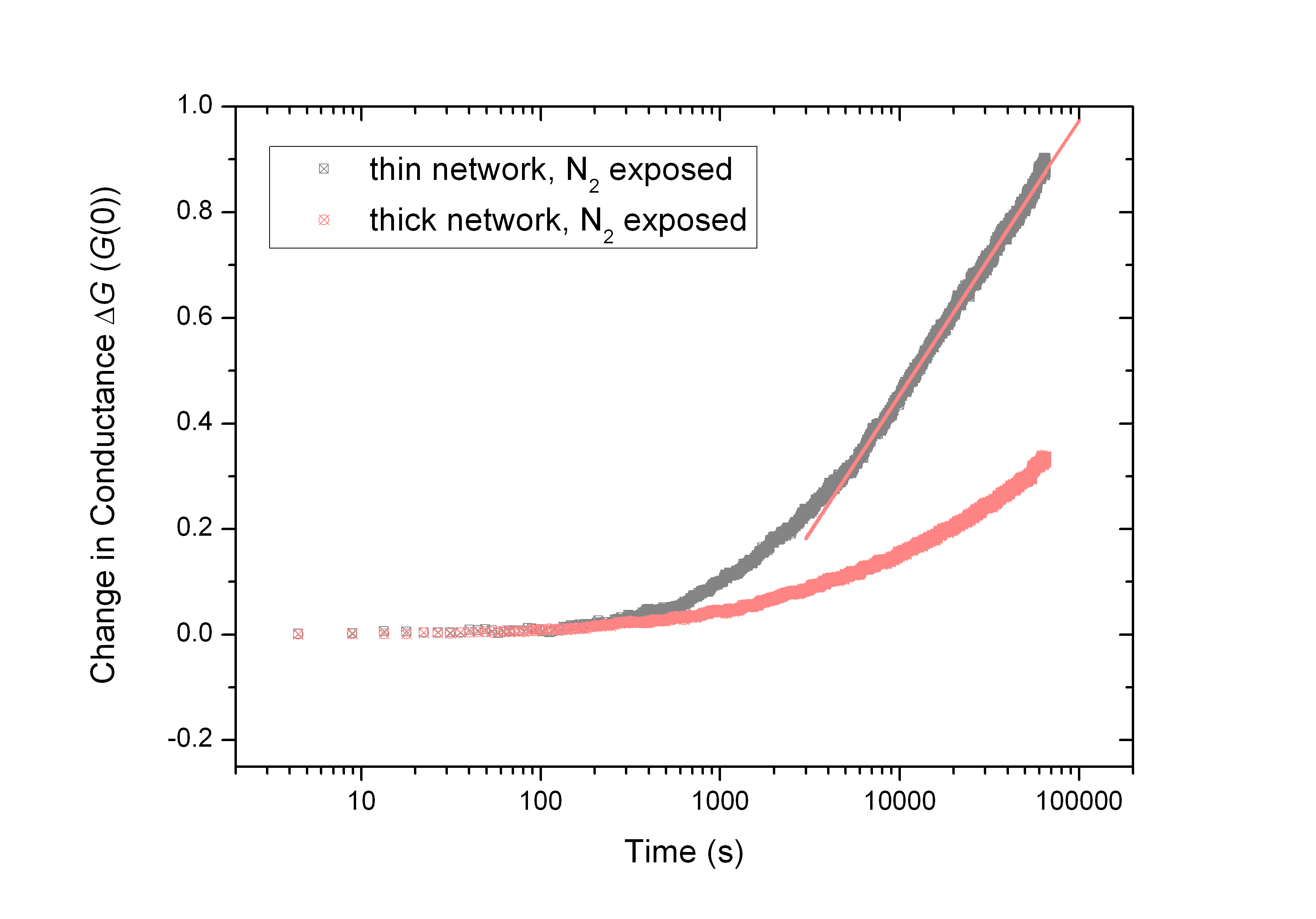}
\caption{Fractional change in conductance $\Delta G = G/G(0)-1$ versus time $t$ in hours for N$_2$ adsorption for thin (${\color{Gray}{\boxtimes}}$) and thick (${\color{Salmon}{\otimes}}$) SWNT networks respectively. Plot on linear-log axis scales one linear region observed for thin network thin network ({\color{Salmon}{\textbf{---$\!$---}}}) and no linear regime for the thick network.}
\label{n2_normal_log}
\end{figure}

The oxygen adsorbed thin and thick SWNT networks depart from Gaussian diffusion at $t \sim 5000$ seconds ($t \sim 1.4$ hours), Figure \ref{SampleA_SampleF_logt}. The nitrogen adsorbed thin SWNT network departs depart from Gaussian diffusive behaviour at $t \sim 400$ seconds, the nitrogen adsorbed thick SWNT network displays a Gaussian diffusion characteristic until the end of the experiment at $t \sim 65000$ seconds, Figure \ref{n2_normal_log}.

Considering the variation in conductance with time to be proportional to the fractional surface coverage of the SWNT network, $\Delta G \propto \Theta$, the thin and thick SWNT networks can be described in this later regime by an Elovich-type adsorption equation

\begin{equation}
\frac{\m{d}\Theta}{\m{d}t} = a^{-b \Theta}\label{SenCond:kine:eq3}
\end{equation}

\noindent where $a$ and $b$ are coefficients and $\Theta$ is the fractional surface coverage. The Elovich isotherm describes adsorption phenomena occurring on surfaces which are energetically heterogenous or on surfaces where the adsorption process induces an energetic heterogeneity \cite{Hubbard2002}. The adsorption kinetics of thin and thick networks during oxygen-adsorption and the thin network during nitrogen adsorption clearly show a linear regime on linear-log axes, Figure \ref{SampleA_SampleF_logt} and Figure \ref{n2_normal_log}, fit by an Elovich type equation

\begin{equation}
\Delta G = \xi \log{(a t)}
\end{equation}

\noindent where $a$ is a constant and $\xi$ is the gradient of the linear fit of $\Delta G$ versus $\log(t)$. The extracted parameters are given in Table \ref{SenCond:kine:DevAF:table2}. There is no Elovich type adsorption behaviour observable for nitrogen adsorption on the thick SWNT network.

\begin{table}\begin{center}%\relsize{-1}
\begin{tabular}{lcccc}
    Sample & $\xi$& Range of $t$ (s)& Duration (h) \\
\hline \hline
thin O$_2$-adsorbed &     $0.34 \pm 0.02$ &    $4900<t<42785$ &     $\sim 11$ \\

thick O$_2$-adsorbed &      $0.14 \pm 0.01$&    $11277<t<83340$ &   $\sim 20$ \\

thin N$_2$-adsorbed & $0.52 \pm 0.01$ & $4702<t <65343$ & $\sim 17$\\
\end{tabular}\end{center}
\caption{Extracted Elovich adsorption parameters from linear fit to Figure \ref{SampleA_SampleF_logt} and Figure \ref{n2_normal_log}.} \label{SenCond:kine:DevAF:table2}
\end{table}

The Elovich isotherm describes any adsorption mechanism that requires an activation energy $\Delta E$ and for which the activation energy increases with surface coverage, $\Theta$, i.e. $\Delta E \propto \Theta$, yielding a general Elovich rate equation of the form

\begin{equation}
\frac{\m{d} \Theta}{\m{d} t} \propto A \exp^{-b \Theta / RT}
\end{equation}

\noindent where $A$ and $b$ are constants and $R$ is the universal gas constant.  The Elovich type adsorption demonstrated by the adsorption kinetics of oxygen adsorbed thin and thick SWNT networks and the nitrogen-adsorbed thin network suggest either preexisting surface heterogeneity or adsorption induced surface heterogeneity. For oxygen adsorption fractional charge-transfer is possible through physisorption \cite{jhi_o2, zhao_adsorption} in addition to chemisorption \cite{topological_o2}, whilst the possibility exists for charge-transfer between the nanotube network and oxygen through electron donation from localised donor states occurring at semiconductor-semiconductor SWNT interfaces \cite{local_DOS}. This process is limited by a potential barrier, $V\s{s}$, formed from the dipole layer created by the redistribution of charge between the O$_2$ molecule and the nanotube network. The increase of this potential barrier with successive adsorption events provides an adsorption induced surface heterogeneity.

The potential barrier, $V\s{s}$, increases with $\Theta$ and the arrival rate, $\m{d}n\s{s}/\m{d}t$, of electrons at the surface, $z = 0$, over $V\s{s}$ is given by \cite{prin_surf_chem}

\begin{equation}
\frac{\m{d}n\s{s}}{\m{d}t} = N \left(\frac{k\s{B}T}{2\pi
m\s{e}}\right)^{1/2} \exp^{-e V\s{s}/k\s{B}T}
\end{equation}

\noindent where $n\s{s}$ is the number of electrons, $N$ is the number of adsorbed species and $m\s{e}$ is the mass of an electron. The rate of $\frac{\m{d}n\s{s}}{\m{d}t}$ is dependent on the specific electronic configuration of the adsorbing gas and adsorbate, therefore the possibility exists to determine the type of gas through the adsorption kinetics. The absence of Elovich adsorption kinetics for the thick network during nitrogen adsorption could be indicative of a reduced charge transfer between nitrogen and the SWNT network, as would be expected for a more metallic SWNT network.

The nanotube network contains many sites for adsorption differing with adsorption energy\cite{zhao_adsorption}. Possible adsorption sites are the nanotube wall, junctions between nanotubes and defects on nanotubes. Recently the electron tunnelling probability between crossed SWNT junctions has been shown to be increased by oxygen and nitrogen adsorption, with a greater increase in tunnelling probability occurring for semiconducting rather than metallic nanotube junctions and with oxygen adsorption increasing the tunnelling probability more than nitrogen adsorption\cite{Mowbray2009}. This provides a convincing origin for the observed adsorption site heterogeneity described by Elovich adsorption. The thick, more metallic, SWNT network displays an Elovich adsorption for oxygen whereas none is visible for nitrogen. This implies that there is a heterogeneity in the energetic distribution of adsorption sites for the adsorption of oxygen on the thick network as a result of perturbation of the intertube tunnelling probability from oxygen adsorption at SWNT junctions. No Elovich type adsorption is observable for the thick SWNT as nitrogen does not significantly alter the intertube electron transmission probability\cite{Mowbray2009}. Thin, more semiconducting, SWNT networks display Elovich adsorption kinetics for both oxygen adsorption and nitrogen adsorption. This is likely a result of the reported enhanced increase in tunnelling probability observable on thin semiconducting SWNT networks for both oxygen and ntirogen\cite{Mowbray2009}.

\section{Conclusion}
We have investigated the adsorption dynamics of oxygen and nitrogen onto thin and thick SWNT networks. Initially the change in conductance with adsorption displays a power-law dependence with time, indicative of Gaussian diffusion of the analyte gas through the networks. Changes in the diffusion parameter are observed during Gaussian diffusion indicating a reduction in the diffusion of gas through the network. At later times an Elovich type diffusion is observed for adsorption of oxygen and nitrogen on thin SWNT networks and oxygen adsorption on thick SWNT network indicating either a pre-existing heterogeneity in adsorption sites or an adsorption induced heterogeneity.

\acknowledgments
C. Morgan acknowledges financial support from the National Physical Laboratory (NPL) and the Engineering and Physical Sciences Research Council (EPSRC).

\section*{References}
\bibliographystyle{unsrt}

\end{document}